# Effective Cost Mechanism for Cloudlet Retransmission and Prioritized VM Scheduling Mechanism over Broker Virtual Machine Communication Framework

Gaurav Raj[1] and Sonika Setia[2]

[1]Asst. Prof. , Lovely Professional University, Phagwara
`er.gaurav.raj@gmail.com`
[2]M. Tech. Student, Lovely Professional University, Phagwara
sonika.setia@hotmail.com

## Abstract

*In current scenario cloud computing is most widely increasing platform for task execution. Lot of research is going on to cut down the cost and execution time. In this paper, we propose an efficient algorithm to have an effective and fast execution of task assigned by the user. We proposed an effective communication framework between broker and virtual machine for assigning the task and fetching the results in optimum time and cost using Broker Virtual Machine Communication Framework (BVCF). We implement it over cloudsim under VM scheduling policies by modification based on Virtual Machine Cost. Scheduling over Virtual Machine as well as over Cloudlets and Retransmission of Cloudlets are the basic building blocks of the proposed work on which the whole architecture is dependent. Execution of cloudlets is being analyzed over Round Robin and FCFS scheduling policy.*

## 1. Introduction

In Current Scenario, with a prior disquisition of the subject, the task is divided and disseminated into same size cloudlets. These Cloudlets as well as Virtual Machines are scheduled according to the First Come First Serve Scheduling Policy i.e. FCFS.

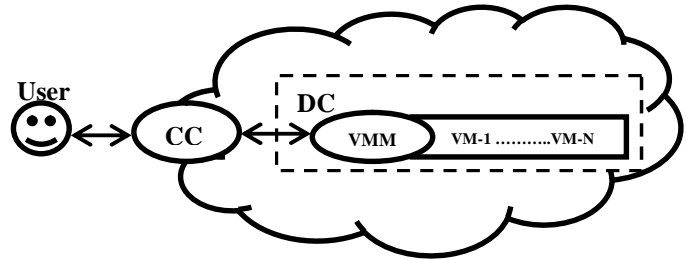

**Fig. 1** General Cloud Computing Scenario

In general Cloud Computing scenario user submits the task to be performed / executed. Cloud Coordinator (CC) divides the task into equal sized cloudlets and passes it to DataCenter (DC). Due to default policy, it takes a lot of time because the cloudlets are processed and emanates one at a time in FCFS manner as and when they reach to VM. In addition to the time consuming factor, the cost factor also acts negatively for this scheduling policy. VM executes the cloudlets present in the queue as they reach the VM's. In a gist, this default policy is extremely Time-Consuming, Cost insensitive and inefficient.

## 2. Related Work

**Gaurav Raj [1],** Author proposed Cost effective Broker Cloud Management where the lowest cost has been calculated in accordance to the communication link. Total cost is calculated on the basis of Hop Count, Bandwidth and Network Delay. An Optimum Route Cost Finder (ORCF) algorithm has been proposed for finding the optimum path on the basis of cost.

**Rodrigo N. Calheiros et al. [2],** Authors gives a generalized framework for modeling and experimentation of the cloud and its working and they called this framework as CloudSim. It gives the benefit to everyone to work and analyse different scenarios over the framework. This reduces the cost as actual implementation requires lot of cost and is almost impossible to implement the desired thing initially at cloud. So, the framework gives the right to simulate various cases before actual implementation.

**Rajkumar Buyya et al. [3],** Authors proposed the market-based resource management that combines the customer service management and the risk management. There is also algorithm and mechanisms for the VM allocation. It also includes market maker, market registry and their services.

**Suraj Pandey et al. [4],** Authors formulate a model to reduce the execution cost of data. It shows that the amount of data being transferred is inversely proportional to the communication cost. Comparison has been made between costs of multiple executions of data with Amazon CloudFront's 'nearest' single data selection. Dividing the data to distributed DC's in proportion to their access cost results into cost savings.

## 3. Proposed Cost Effective Framework with Scheduling Algorithm

Our main aim is to schedule the cloudlets as well as VM so that it provides more efficient and effective results.

In BVCF, User assigns the task to CC through Broker which acts as intermediate. CC divides the task into same sized cloudlets and sends it further to DC. Every DC has its own VM Manager (VMM) which manages the Virtual Machine Pool (VMP). VMM requests for the resources to the Resource Provisioner (RsP) and it then further asks for access permission from Resource Provider (RP) and responds back to the VMM. After the permissions being granted by RP, RsP provide resources, based on minimum cost of resources, to VMM for creating VM. The cost is calculated using the below formula:

$$T_C = \quad H_C + \quad N_D + \quad B + \quad S_C$$

Here, $H_C$ is Hop Count, $N_D$ is Network Delay, B is Bandwidth and $S_C$ is Security Cost . And , , and are variables [1].

Analysing the total cost, VMM prioritizes the different VM's and chooses the VM with the highest priority or lowest cost. The VMM sends the cloudlet ID list consists of all the ID's of the cloudlets received from the CC. Then, VMM sends actual cloudlets whose ID is being checked by the VM and matched with the ID in cloudlet ID list. After receiving the accurate cloudlet VM sends the Acknowledgement otherwise the retransmission message is sent back. After this, the cloudlet is executed by two type of scheduling policies i.e.

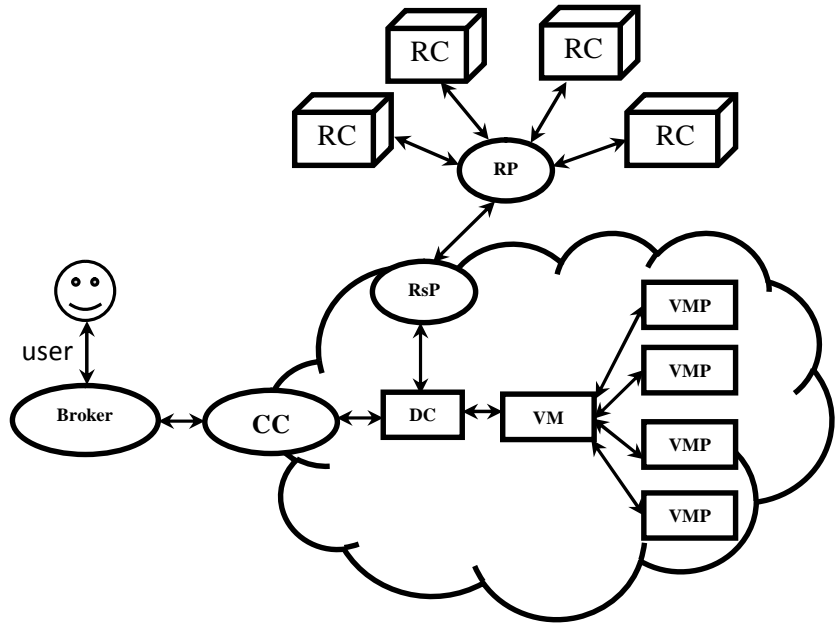

**Fig. 2** Broker Virtual Machine Communication Framework (BVCF)

FCFS over cloudlets – The cloudlets will execute in the sequence as they are entering the VM. Round-Robin over cloudlets – Initially the Time Quantum is assigned by the User and each and every cloudlet executes for that time quantum and left cloudlets will be executed afterwards for that time quantum only.

## 3.1 Proposed Scheduling and Retransmission Algorithm:

The main algorithm for the proposed architecture is divided into further three sub algorithms which are:

   i.  **Cost Computation Algorithm** – Computation of Cost
   ii. **Cloudlet Retransmission and Prioritized VM Algorithm** – Prioritizing the VM's and Sending and Retransmitting the cloudlets.
   iii. **Cloudlet Scheduling Algorithm** – Scheduling the Cloudlets.

### 3.1.1 Algorithm-1:

It is named as Cost Computation Algorithm. Its basic purpose is to compute the cost at the time of Virtual Machine creation.

---

**Algorithm-1** Cost Computation Algorithm

**Step-1:**  User assigns task to Cloud Co-ordinator.

**Step-2:**  Cloud Co-ordinator divides the task into cloudlets.

$$\text{Cloudlet\_list} \quad \leftarrow \quad \text{task}$$

**Step-3:**  Virtual Machine Manager creates the Virtual Machine.

**Step-4:**  At the time of creation, Resource Provider allocates the resource to Virtual Machine.

$$\text{Virtual Machine} \quad \leftarrow \quad \text{Resource Provider [Resources]}$$

**Step-5:**  Total Cost is calculated keeping in mind four factors i.e. Hop count, Bandwidth, Network

Delay and Security Cost.

$$T_C = \quad H_C + \quad N_D + \quad B + \quad S_C$$

---

### 3.1.2 Algorithm-2:

It is named as Cloudlet Retransmission and Prioritized VM Algorithm. Its purpose is to acknowledge or retransmit the cloudlets which are received by the VM.

---

**Algorithm-2** Cloudlet Retransmission and Prioritized VM Algorithm

**Step-1:** VMM sets priority of VM's according to $T_C$

*Set Priority* of VM_list[i] using $VM_{Priority} \propto 1/T_C$

**Prioritized VM list = VM list**

**Step-2:** VMM arrange / sorts the cloudlet list and store them as cloudlet_ID_list.

**Step 3:** Sends the *cloudlet_ID_list* to VM

**Step-3:** VMM sends *cloudlet_ID_list* to VM as *Predicted Sequence List.*

**Predicted Sequence List** ← **Cloudlet_ID_List**

**Step-4:** VMM sends **actual cloudlets** to VM.

While **cloudlet_list == Empty**

oudlet_list[i] = VMM_cloudlet_list[i]

**Step-5:** *If* **actual_cloudlet[ID] = Predicted Sequence List[ID]**

  *Then,* continue till **(received_cloudlet_list).Length == 10** and

  send **acknowledgement[receiving_cloudlet_list[ID]].**

ait[unit_time] and

    **Retransmit[cloudlet]** and **counter++.**

    Here, **counter = received_cloudlet_list[ID].**

---

### 3.1.3 Algorithm-3

It is named as Cloudlet Scheduling Algorithm. Its purpose is to schedule the cloudlets according to FCFS and Round Robin Scheduling policy.

---

**Algorithm-3** Cloudlet Scheduling Algorithm

**Step-1:** Choose the type of Scheduling

   *1. Priority Based First Come First Serve Scheduling*

   *2. Priority Based Round Robin Scheduling*

  User enters the choice.

    **Choice == 1 || 2**

**Step-2:** If choice is 1.

  Then

    goto **Label 1**.

  If choice is 2.

  Then

    goto **Label 2**.

**Step-3:** Executed cloudlets are returned to Cloud Co-ordinator.

   **CC [cloudlet_list]** ← **VMM [executed_cloudlet_list]**

**Step-4:** Cloud Co-ordinator combines all the cloudlets to form task.

   **Combine [cloudlets]**

**Step-5:** Executed Task returned back to User by Cloud Co-ordinator.

   **User[executed task]** ← **combine[cloudlets]**

---

FCFS and Round Robin Scheduling are shown in Label 1 and Label 2 respectively.

---

**Label 1:** FCFS Scheduling

      For every cloudlets received

                  **Executed_list[cloudlets]** ← **Execution [Actual_cloudlet]**

                  **VMM[executed_cloudlet_list]** ← **Executed_list[cloudlets]**

      goto Step3

---

**Label 2:** Round Robin Scheduling

      [*initialize*] time quantum as **tq = 10**

      Repeat Steps till **(actual_cloudlet_list).Length == NULL**

      {

            **Execute[actual_cloudlet]** till **tq**

            **Executed_list[cloudlets]** ← **Execution[Actual_cloudlet]**

            **actual_cloudlet** ← **cloudlet_list[next_cloudlet]**

      }

      **VMM [executed_cloudlet_list]** ← **Executed_list [cloudlets]**

      goto Step3

---

### 3.2 Sequence Diagram for the Procedure:

1. User *assigns* the *task* to CC.
2. CC *divides* the assigned task *into* same sized *cloudlets.*
3. CC *sends* the *cloudlets* to VMM.
4. VMM *sends* the *list* of the *needed resources* to the RsP.
5. RsP *requests* the *resources* from RP.
6. RP provides *access* to use the resources.
7. RsP *grants* the *access* to VMM.
8. VMM *creates* the VM using the resources.
9. *Total Cost* of VM is calculated by using factors like *Hop Count, Network Delay, Bandwidth and Security Cost.*
10. VMM *prioritizes* the VMs according to the *cost factor* and selects the VM with *highest priority or lowest cost.*
11. VMM *sends* the *cloudlet ID list* to VM.
12. VMM *sends* the *actual cloudlet* to VM.
13. VM *matches* the *ID of actual cloudlet* with the *sequence of cloudlet ID list.*
14. If both *ID matches* then, VM *sends* the *acknowledgement* to VMM. *Otherwise*, VM sends *Retransmit message.*

**15.** Request for *execution* of the *cloudlet* is sent to the VM by VMM.

**16.** *Cloudlet scheduling* is done by VM according to *FCFS and Round Robin*.

**17.** VM *sends* the *executed cloudlets* to the VMM.

**18.** VMM further *passes* the *executed cloudlets* to CC.

**19.** CC *combines* all the *executed cloudlets together* to form the task.

**20.** CC *sends* the *executed task* back to the user.

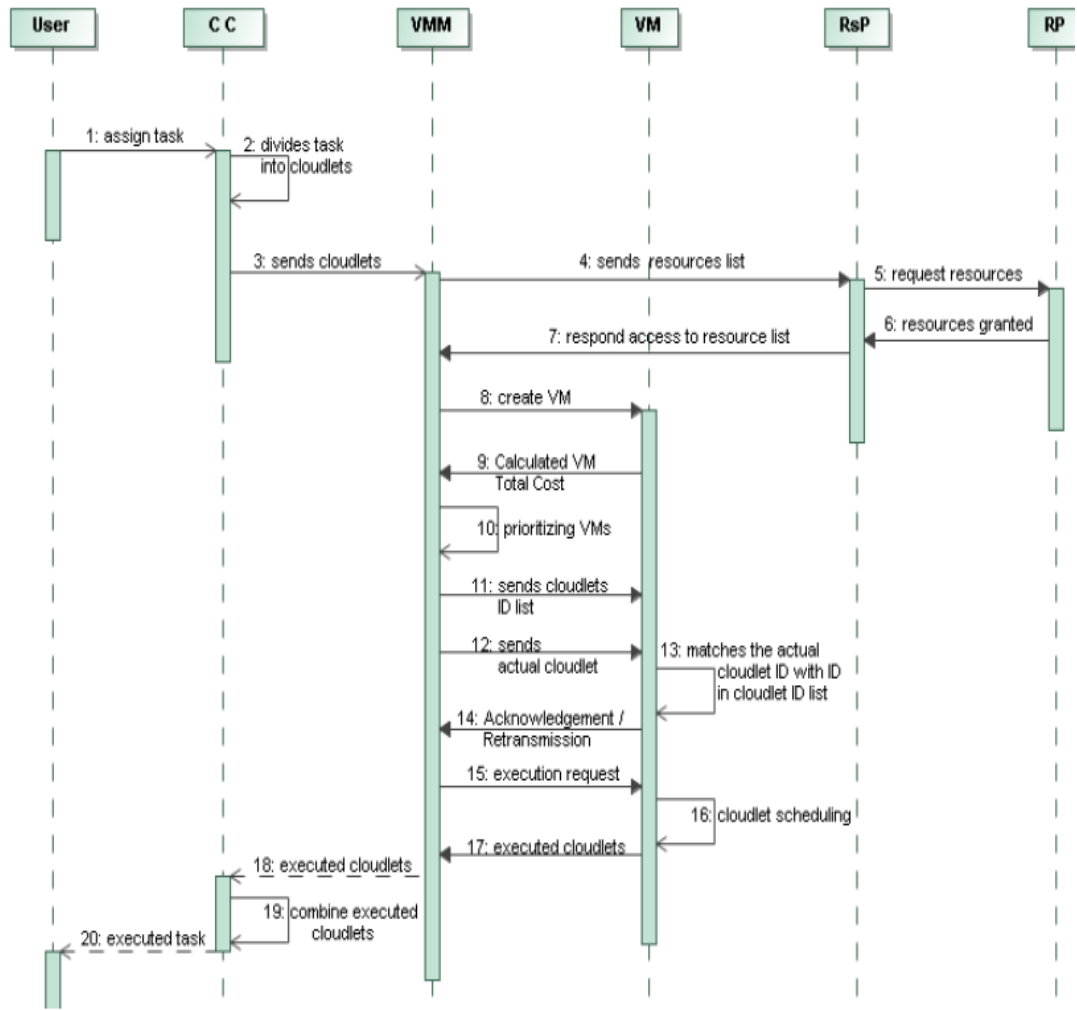

**Fig. 3** Sequence Diagram of work flows for the task execution

In Fig, 3, there are six actors i.e. User, Cloud Co-ordinator, VM Manager, Virtual Machine, Resource Provisioner and Resource Provider. The whole process takes place in eighteen steps showing which actor is performing which task and how the assigned task is executed. All the steps or method shown in the sequence diagram are explained below as:

## 3.3 Use Case Diagram of the Procedure:

There are six actors in the above Use Case Diagram:
  a) User
  b) Cloud Coordinator (CC)
  c) Virtual Machine Manager or VM Manager (VMM)
  d) Virtual Machine (VM)
  e) Resource Provisioner (RsP)
  f) Resource Provider or Resource Owner(RP)

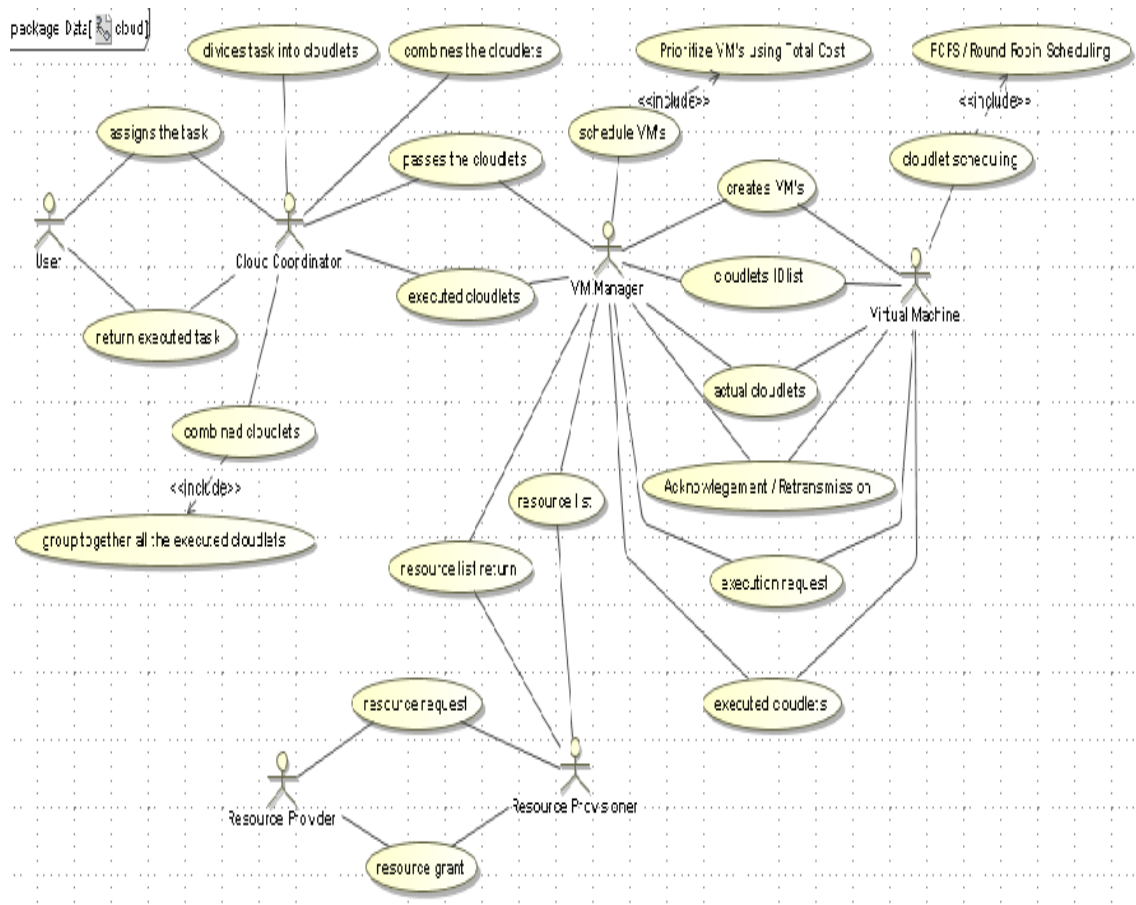

**Fig. 4** Use Case Diagram of BVCF, showing all the actors and the work performed by the actors.

Cloud Coordinator itself divides or combines the task to cloudlets so it is not connected to other actor.

Similarly, VM Scheduling is done by VM Manager and Cloudlet Scheduling is done in Virtual Machine.

48

Include Relationship is shown if the use case further have some use case. Like in this diagram, Schedule VM includes the Prioritizing VM's using Total Cost. This means schedule VM depends upon the prioritization of VM using Total Cost.

## 4. Results

After implementing the proposed algorithms over BVCF, results for computation of cost are obtained which further helps in the study of comparison between the FCFS and Round Robin Scheduling policy.

The whole procedure is being done with the two different scheduling policies and the cost of both the procedures is shown in Fig 5.

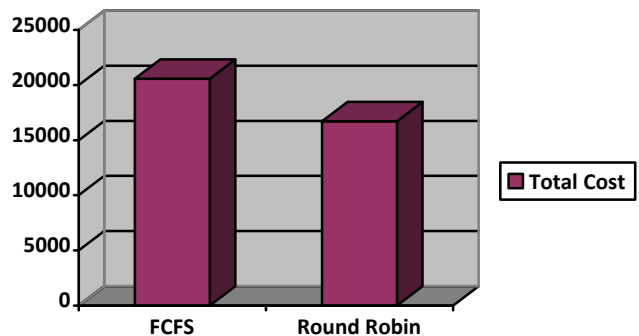

**Fig. 5** Comparison between FCFS and Round Robin

Same numbers of virtual machines are considered while calculating total cost for both cases i.e. FCFS and Round Robin. The location of user i.e. from where the user is demanding the services is also considered same for both the cases. The total cost for executing all cloudlets is more in case of FCFS scheduling policies. And due to the Acknowledgment and Retransmission criteria, all cloudlets must run. This is why no cloudlet results in paused mode, or stopped mode.

*Fig. 5* shows that the total cost of all cloudlets being executed with the help of Round Robin is comparatively lower than that of being executed by FCFS scheduling policy. So, the proposed algorithm succeeds in reducing the cost.

## 5. Conclusion

Cost and time are always the grassroots that come into mind either before buying or selling any services. In this paper, we have proposed an algorithm for optimized Cost of VM, prioritizing the VM and cloudlet scheduling through FCFS and Round Robin. BVCF explains the whole proposed scenario in which the execution of task is performed. The proposed algorithm helps in the Fast Execution due to Round Robin Scheduling Policy applied on the equally sized cloudlets. All cloudlets will execute as after each and every successfully received cloudlet VM sends the acknowledgement and for the unsuccessful cloudlets sends the retransmit message. It also results into Lower Cost as the VM's are prioritized according to its Cost only.

## 6. Future Work

Cloud Computing is a perpetual research as all its aspects haven't unfolded yet. BVCF is just a framework where most of the mechanisms can be implemented so as to analyze or optimize the performance of any of the midget area. Security is the major issue in every area and we can also implement various security mechanisms over BVCF. Resources can also be allocated accordingly to reduce the cost. VM's images can be saved in the database so that if the same task is being assigned by the user then we can allocate the same VM with optimal resources. Cost of VM can also include certain other factors on which it is dependent.

## Authors

1. Sonika Setia
Student of M.Tech in Information Technology from Lovely Professional University, Phagwara.

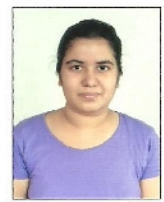

2. Gaurav Raj

Asst. Prof. at Lovely Professional University, Phagwara and has done M.Tech from Motilal Nehru National Institute of Technology, Allahabad.

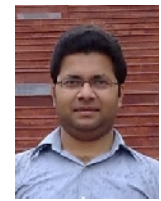